\documentclass[12pt]{article}
\usepackage{amssymb}

\newcommand{\nc}{\newcommand}
\nc{\la}{\lambda} \nc{\alf}{\alpha}
\nc{\tht}{\theta}  \nc{\be}{\beta}  \nc{\eps}{\epsilon}
\nc{\ga}{\gamma}  \nc{\De}{\Delta}  \nc{\Ga}{\Gamma}  \nc{\vphi}{\varphi}
\nc{\de}{\delta} \nc{\si}{\sigma}  \nc{\ka}{\kappa}
\nc{\om}{\omega}  \nc{\Om}{\Omega}  \nc{\La}{\Lambda}
\nc{\qq}{\quad\quad}  \nc{\nf}{\infty}   \nc{\dl}{\mathop{\smash{\cal L}}}
\nc{\ra}{\rightarrow}  \nc{\ol}{\overline}  \nc{\und}{\underline}
\nc{\beq}{\begin{equation}}   \nc{\eeq}{\end{equation}}
\nc{\beqa}{\begin{eqnarray}}  \nc{\eeqa}{\end{eqnarray}}
\nc{\nin}{\noindent}          \nc{\pt}{\partial}
\nc{\dst}{\displaystyle}      \nc{\nnb}{\nonumber}
\nc{\bs}{\backslash}          \nc{\mb}{\mathbb} 
\nc{\dg}{\dagger}   \nc{\wh}{\widehat}  \nc{\wti}{\widetilde}
\nc{\PT}{\Pi_{\tht}}  \nc{\PF}{\Pi_{\phi}}       \nc{\R}{\mb R}

\newcounter{muni}
\newenvironment{remunerate}{\begin{list}{{\rm \arabic{muni}.}}
{\usecounter{muni}
\setlength{\leftmargin}{0pt}\setlength{\itemindent}{38pt}}}{\end{list}}

\nc{\brm}{\begin{remunerate}}   \nc{\erm}{\end{remunerate}}
\nc{\barr}{\begin{array}}   \nc{\earr}{\end{array}}

\nc{\stg}{\mathop{\smash{*}}}    \nc{\st}{\mathop{\smash{\delta}}}
 
\nc{\mtvb}{\mathversion{bold}}   \nc{\mtvn}{\mathversion{normal}}

\begin{document}

\begin{titlepage}

\[  \]
\centerline{\Large\bf  One loop renormalizability of }

\vspace{4mm}
\centerline{\Large\bf the Poisson-Lie sigma models}

\vskip 2.0truecm
\centerline{\large\bf Galliano VALENT${}^{\;\dagger\;}\qq\qq$ Ctirad KLIM\v{C}\'IK${}^{\;\ddagger\;}$}

\vskip 1.0truecm
\centerline{\large\bf  Romain SQUELLARI${}^{\; *\;}$}

\vskip 2.0truecm
\centerline{${}^{\dagger\,*}$ \it Laboratoire de Physique Th\'eorique et des
Hautes Energies}
\centerline{\it CNRS, Unit\'e associ\'ee URA 280}
\centerline{\it 2 Place Jussieu, F-75251 Paris Cedex 05, France}
\nopagebreak

\vskip 0.5truecm
\centerline{${}^{\ddagger\,*}$ \it Institut de Math\'ematiques de Luminy}
\centerline{\it CNRS UPR 9016}
\centerline{\it Case 907, 163 Avenue de Luminy}
\centerline{\it 13288 Marseille Cedex 9, France}
\nopagebreak

\vskip 2.5truecm

\begin{abstract}
We present the proof of the one loop renormalizability in the strict field theoretic sense 
of the Poisson-Lie $\si$-models. The result is valid for any Drinfeld double and it
relies solely on the Poisson-Lie structure encoded in the target manifold.

\end{abstract}

\end{titlepage}

\title{\bf Poisson Lie models: one-loop renormalizability}

\nc{\Tt}{\wti{T}} \nc{\tA}{\wti{A}} \nc{\tB}{\wti{B}}  \nc{\tG}{\wti{G}}  \nc{\tPi}{\wti{\Pi}}
\nc{\tf}{\tilde{f}} \nc{\tg}{\tilde{g}}  \nc{\lan}{\langle}  \nc{\ran}{\rangle}

\section{Introduction}
At the classical level, the $T$-duality  \cite{KY,SS,FJ,AAL,KS1}  is  an isomorphism between 
nonlinear $\sigma$-models  with geometrically different target manifolds. Or, speaking 
more precisely, it  is the  symplectomorphism between the phase spaces of the 
$\sigma$-models that transforms the Hamiltonian of one model into the Hamiltonian 
of its dual: it relates  right-invariant geometries on a Lie group target $G$ to certain 
geometries on the dual space ${\cal G}^*$ of the Lie algebra ${\cal G}:=Lie(G)$.  
At the quantum level, the status of $T$-duality is much more subtle, mainly 
because the geometries  of the target manifolds generically receive quantum corrections. 
However, in the particular case of the  so called conformal field theories, 
the $\sigma$-model geometries do not receive such corrections and the $T$-duality can 
be understood as a unitary transformation relating the spectra and the correlation 
functions of the {\it quantum}  $\sigma$-models.

Outside from the conformal points, the original and the dual geometry acquire a dependence 
on the cut-off. However, if this dependence  is such that the geometries  remain $T$-dual 
to each other for every value of the cut-off then $T$-duality is preserved also at the 
quantum level. 

This equivalence problem can be analyzed either in the 
stringy framework where the $\si$-model interacts with matter and gravity or for the 
pure $\si$-model, in which case one has first to prove renormalizability in the strict 
field theoretic sense and then study the equivalence problem. 
In the first setting a general proof that the one loop equivalence works 
was given in \cite{Ty} provided that the structure constants of ${\cal G}$ are traceless. 
In the second setting, the one-loop renormalizability and the quantum equivalence 
work without any restriction \cite{CV}.

The standard non-Abelian $T$-duality has a nontrivial generalization referred to as the 
Poisson-Lie $T$-duality \cite{KS1,KS2}.  Both the original and the dual geometries of the
Poisson-Lie $T$-dualizable $\sigma$-models are derived from  the so called 
Drinfeld double which is a Lie group equipped with some additional structure (cf. Section 2). 
The one loop equivalence in the stringy framework was shown to work in \cite{BM} provided that all 
the structure constants of the double are traceless. In the restricted field theoretic 
framework, the one loop renormalizability and compatibility with the renormalization 
were so far established only for very few special pairs of the mutually dual $\si$-models 
\cite {Sf1,KV}. Those special models live on target manifolds with small dimensions 
and their geometries can be explicitly calculated in suitable coordinates. At a first sight, 
it appears to be too difficult a task to deepen this result. Indeed, there are very numerous different Drinfeld doubles (they are far from being all classified) and  even  for the known 
Drinfeld doubles (like, for example, simple complex groups), the explicit characterization 
of the dualizable geometries in terms of  coordinates on group manifolds becomes forbiddingly 
complicated for group targets with dimension 4 and more. In spite of  all those obstacles,  
in this paper we do establish the one loop renormalizability of the Poisson-Lie $\si$-models 
in full generality, i.e. for an arbitrary Drinfeld double. We succeed to obtain
this general result mainly because we can express the Ricci tensor in a tractable form 
just using frames rather than introducing coordinates on the target manifold. 

To tell the truth, at the very beginning of our calculation,  we did not aim as much as 
we have eventually obtained. We simply wanted  to identify which additional properties 
a Drinfeld double must possess in order to ensure the one loop renormalizability. 
To our  satisfaction, we find that  no such additional properties are needed. In fact, 
we have been impressed how naturally just the basic  Poisson-Lie 
structure (and nothing else) has been sufficient to tame the ultraviolet divergences.
 
The plan of the paper is as follows. In Section 2, we review the concept of the 
Drinfeld double and we show how the target geometries are associated to it. In Section 3, 
the riemannian geometry with torsion is developed and leads to a nice form of the 
Ricci tensor. This result is then used, in Section 4, to establish the 
one-loop renormalizability of the Poisson-Lie $\si$-model.

\section{Drinfeld doubles and Poisson-Lie $T$-duality}
 
Consider  a basis  $T_a$, $a=1,...,n$  in a  vector space  ${\cal G}$ and the dual basis $\Tt_a$ 
in the dual space $\wti{\cal G}$. Equip both ${\cal G}$ and $\wti{\cal G}$ with  Lie algebra structures
\[[T_a,T_b]=f_{ab}^s\,T_s,\qq [\Tt^a,\Tt^b]=\tilde{f}^{ab}_s\,\Tt^s.\]
One says that the Lie algebras ${\cal G}$ and $\wti{\cal G}$ are compatible  if the following brackets 
\[  [T_a,T_b]_{\cal D}=f_{ab}^s\,T_s,\qq [\Tt^a,\Tt^b]_{\cal D}=\tilde{f}^{ab}_s\,\Tt^s \qq 
[T_a,\Tt^b]_{\cal D}=\tilde{f}^{bs}_a\,T_s-f^b_{as}\,\Tt^s\]
define a  Lie algebra structure on the direct sum  vector space 
${\cal D}:={\cal G}\stackrel{.}{+}\wti{\cal G}$. In this case, we say that the Lie algebra 
${\cal D}$ is the {\it Drinfeld double} of ${\cal G}$ 
(or, equivalently, of $\wti{\cal G}$). Note that  the Drinfeld double  ${\cal D}$ comes equipped 
with an $Ad$-invariant bilinear form $\lan .,.\ran_{\cal D}$ defined as
\beq \lan T_a,\Tt^b \ran_{\cal D}=\delta_a^b,\quad 
\lan T_a,T_b \ran_{\cal D}=\lan \Tt^a,\Tt^b \ran_{\cal D}=0.\label{form}\eeq
Consider the connected and simply connected group $D$ the Lie algebra of which is ${\cal D}$ and 
the subgroups  $G$ and $\wti{G}$ of $D$  corresponding to the subalgebras ${\cal G}\subset {\cal D}$ 
and $\wti{\cal G}\subset{\cal D}$, respectively.  The  group $D$ is called the Drinfeld double 
of $G$ (or of $\wti{G}$).
 
In what follows, we shall be often using the matrices of the adjoint action 
of $D$ on ${\cal D}$ in the basis $T_a,\Tt^b$:
\[Ad_g\,\Tt^a\equiv l^{-1}\,T\, l=B^{as}(g)Ts+A^a_s(g^{-1})\Tt^s,\quad g\in G,\]  \[Ad_{\tg}\,T_a=\tB_{as}(\tg)\Tt^s+\tA_a^s(\tg^{-1}) T_s, \quad \ \tg\in\tG.\]
We can write those matrices also in terms of the bilinear form (\ref{form}) as
\[A_a^b(g):=\lan Ad_g\,T_a,\Tt^b \ran,  \qq B^{ab}(g):=\lan Ad_g\,\Tt^a,\Tt^b\ran;\]
\[\tA_a^b(\tg):=\lan T_a, Ad_{\tg}\,\Tt^b\ran,  \qq \tB_{ab}(\tg):=\lan Ad_{\tg}\,T_a,T_b\ran.\]
It turns out that the algebraic  structures that we have introduced so far can be used to define 
certain Poisson brackets  of functions on the groups $G$ and $\tG$:
\[\{f_1,f_2\}(g):=\Pi^{ab}(g)\nabla_af_1(g) \nabla_bf_2(g), \qq 
\{f_1,f_2\}^*(\tg):=\tPi_{ab}(\tg)\nabla^a \tf_1(\tg) \nabla^bf^*_2(\tg),\]
where the right-invariant vector fields $\nabla_a$ and $\wti{\nabla}^a$  on $G$ and $\wti{G}$ are defined by
\beq\label{nabla}
\nabla_af(g):=\frac{d}{ds}f(e^{sT_a}g)\Big|_{s=0}, \qq   \wti{\nabla}^a\tf(\tg):=\frac{d}{ds}\tf(e^{s\Tt^a}\tg)\Big|_{s=0}\eeq
and the antisymmetric matrix-valued functions  $\Pi^{ab}(g)$ and $\tPi_{ab}(\tg)$ are given by
\[ \Pi^{ab}(g):=-B(g)_{as}A_s^b(g^{-1}), \quad g\in G;\qq  
\tPi_{ab}(\tg):=-\tB_{as}(\tg)\tA^s_b(\tg^{-1}), \quad \tg\in \tG.\]
The Poisson  structures on $G$ and $\tG$ turn out to satisfy the so-called cocycle conditions 
in the ${\cal G}\wedge {\cal G}$ and $\wti{\cal G}\wedge \wti{\cal G}$-valued group cohomologies 
of $G$ and $\tG$, respectively:
\beq 
\Pi(hg)=\Pi(h)+Ad_h\Pi(g), \quad g,h\in G; \quad  \tPi(\tilde{h} \tg)=
\tPi(\tilde{h})+Ad_{\tilde{h}}\tPi(\tg),\quad \tg,\tilde{h}\in \tG.\label{coc}\eeq 
Here we have set
\[\Pi(g):=\Pi^{ab}(g)T_a\otimes T_b, \qq \tPi(\tg):=\tPi_{ab}(\tg)\Tt^a\otimes \Tt^b.\]
 
Consider an invertible matrix $M^{ab}$ and the  maps $E_M:G\to{\cal G}\otimes{\cal G}$ and $\wti{E}_M:\wti{\cal G}\to\wti{\cal G}\otimes\wti{\cal G}$ defined by
\beq 
E_M(g):=E_M(g)^{ab}T_a\otimes T_b; \qq \wti{E}_M(\tg):=\wti{E}_{M}(\tg)_{ab}\Tt^a\otimes \Tt^b,
\eeq
\beq   
E_M(g)^{ab}:=M^{ab}+\Pi^{ab}(g);  \qq \wti{E}_{M}(\tg)_{ab}:=(M^{-1})_{ab}+\tPi_{ab}(\tg).
\eeq
The Poisson-Lie $T$-duality then establishes the isomorphism between $\sigma$-models
living on the targets $G$ and $\tG$. These  models  are completely specified by the  maps  $E_M$ and
$\wti{E}_M$. Their respective field configurations are smooth maps  $g:W\to G$  and $\tg:W\to \tG$, 
where $W$ is the   two-dimensional world-sheet, and  their respective dynamics are 
determined  by the least action principles:
\beq\label{orig}  
S(g)=\int_W \Big( E_M(g)^{-1}, R(g)_+\otimes R(g)_-\Big) d\xi^+ d\xi^- ;\eeq
\beq\label{dual}  
\wti{S}(\tg)=\int_W \Big( \wti{E}_M(\tg)^{-1} ,R(\tg)_+\otimes R(\tg)_-\Big) d\xi^+ d\xi^- .\eeq
Here $(.,.)$ is the duality pairing between ${\cal G}\otimes {\cal G}$ and 
$\wti{\cal G}\otimes\wti{\cal G}$, $\xi^+,\xi^-$  are the light-cone coordinates on $W$, 
$R(g)_+d\xi^+ + R(g)_-d\xi^-$ denotes  the pull-back of the right-invariant Maurer-Cartan 
form on $G$ by the map $g:W\to G$ and $E_M^{-1}:G\to \wti{\cal G}\otimes\wti{\cal G}$ is 
inverse  to $E_M$, i.e. 
\[E_M(g)\Big(., E_M(g)^{-1}(.,u)\Big)=u, \qquad \forall g\in G, \quad  \forall u\in{\cal G}.\]
Similarly, $R(\tg)_+d\xi^+ + R(\tg)_-d\xi^-$ denotes  the pull-back of the right-invariant 
Maurer-Cartan form on $\tG$ by the map $\tg:W\to \tG$ and 
$\wti{E}_M^{-1}:\tG\to {\cal G}\otimes{\cal G}$ is inverse  to $\wti{E}_M$.
 
If we introduce some  local coordinates $X^\mu$ e.g. on the target $G$, any field configuration $g$ can be locally viewed as a collection of real functions $X^\mu(\xi^+,\xi^-)$  in terms of which the action (\ref{orig}) can be locally rewritten  as
\beq\label{deom} 
S=\int_W (g_{\mu\nu}(X)+h_{\mu\nu}(X))\partial_{\xi^+}X^\mu\partial_{\xi^-}X^\nu d\xi^+ d\xi^- .\eeq
Here  $g_{\mu\nu}$ is a  symmetric tensor interpreted as a metric on $G$ and  $h_{\mu\nu}$ is an antisymmetric tensor interpreted as a torsion potential on $G$.  Thus we see that any choice of the map $E_M:G\to {\cal G}\otimes{\cal G}$ defines an $M$-dependent  {\it geometry} on $G$.
 
We note one crucial fact: the moduli space of the dual pairs of the $\sigma$-models 
(\ref{orig}), (\ref{dual}) associated to a given Drinfeld double $D$ is {\it finite-dimensional} 
since it is parametrized by the invertible matrices $M$. If we wish that the quantum corrections 
do not spoil the $T$-dualizability, all  ultraviolet divergences must be eliminated just by a suitable cut-off dependence of the matrix $M$.  In other words, if we interpret the entries of the matrix $M$ as the coupling constants, we simply require the renormalizability of the model  (\ref{orig}) in the standard field theoretic sense of this term.  As we shall see in the next section, this is precisely what happens.
     
The matrices $\Pi^{ij}(g)$ and $\tPi_{ij}(\tg)$ appear explicitly in the $\sigma$-model 
Lagrangians on the targets $G$ and $\tG$, it is therefore obvious that the countertems needed to cancel the ultraviolet divergences must depend algebro-differentially on them. 
Actually, if $\Pi$ and $\tPi$ were generic matrix-valued functions on $G$ and $\tG$,
 the dependence of the counterterms on them would be too complicated to ensure renormalizability, However, due to  their special definitions, $\Pi$ and $\tPi$  satisfy two crucial identities 
which, quite remarkably,  are sufficient to disentangle the counterterm structure and ensure renormalizability. They were first derived in \cite{Sf2} in a completely algebraic way, but it is perhaps
more insightful to understand these relations as resulting from the Poisson-Lie 
geometry, as shown in \cite{Kl2}. The first of those identities is the direct consequence of the cocycle condition (\ref{coc}) for $h$ and $\tilde{h}$ respectively  close to the group units of $G$ and $\tG$ .  It reads
\beq\label{idPi}
\nabla_c\Pi^{ab}(g)=\tf^{ab}_c-f^a_{cs}\Pi^{bs}(g)+f^b_{cs}\Pi^{as}(g)\eeq 
\beq\label{idtPi}
\wti{\nabla}^c\tPi_{ab}(\tg)=\ f^c_{ab}-\tf^{cs}_a\tPi_{bs}(\tg) 
+\tf^{cs}_a\tPi^{as}(\tg)\eeq
The second identity is nothing but the Jacobi identity for the Poisson brackets on $G$ and on $\tG$:
\beq\label{idQ1}
\Pi^{sa}(g)\nabla_s\Pi^{bc}(g)
+f^a_{st}\Pi^{bs}(g)\Pi^{ct}(g) +cp(a,b,c)=0\eeq
\beq\label{idQt1}
\tPi^{sa}(\tg)\wti{\nabla}^s\tPi_{bc}(\tg)
+\tf_a^{st}\,\tPi^{bs}(\tg)\tPi^{ct}(\tg)+cp(a,b,c)=0\eeq
where $cp(a,b,c)$ means circular permutation of the indices involved. As a consequence of 
(\ref{idPi}), (\ref{idtPi}), (\ref{idQ1}) and (\ref{idQt1}), we obtain
\beq\label{idQ2}
 f^a_{st}\Pi^{bs}(g)\Pi^{ct}(g)+\tf^{ab}_s\Pi^{cs}(g)+cp(a,b,c)=0.\eeq
\beq\label{idQt2}
\tf_a^{st}\tPi_{bs}(\tg)\tPi_{ct}(\tg)+f_{ab}^s\tPi^{cs}(\tg)+cp(a,b,c)=0.\eeq

\section{The geometry of Poisson-Lie $\si$-models}
\subsection{Geometry with torsion}
The form (\ref{orig}) of the model gives a prominent role to the right-invariant frames:
\beq
dg\,g^{-1}=R^a(g)\,T_a,\qq\qq dR^a(g)=\frac 12\,f^a_{bc}\,R^b(g)\wedge R^c(g),\eeq
leading us to introduce
\beq\label{metric}
G_{ab}\,R^a(g)\,R^b(g),\qq\qq G_{ab}=g_{(ab)}+h_{[ab]},\eeq
where $g$ will be a riemannian metric and $h$ the torsion potential. >From $h$ we get the torsion 3-form according to
\beq
H=\frac 12\ h_{ab}\,R^a\wedge\,R^b\quad\Rightarrow\quad  
T=dH=\frac 1{3!}\,T_{abc}\,R^a\wedge R^b\wedge R^c,\eeq
and we will need also
\beq
T^a=\frac 12\,T^a_{bc}\,R^b\wedge R^c,\qq T^a_{bc}=g^{as}\,T_{sbc}.\eeq 

Putting coordinates $\{X^{\mu}\}$ on the group we recover the metric and the torsion 
potential defined in (\ref{deom}) by
\beq
g_{\mu\nu}=g_{ab}\,R^a_{\mu}\,R^b_{\nu},\qq\quad h_{\mu\nu}=h_{ab}\,R^a_{\mu}\,R^b_{\nu},\eeq
and the symmetric connection $\nabla$ defined as usual by
\beq
\nabla_{\mu}v_{\nu}=\pt_{\mu}v_{\nu}-\ga^{\si}_{\mu\nu}v_{\si},\qq 
\ga^{\alf}_{\mu\nu}=\frac 12\,g^{\alf\si}\Big(\pt_{\mu}g_{\nu\si}+\pt_{\nu}g_{\mu\si}-\pt_{\si}g_{\mu\nu}\Big).
\eeq
One can define two connections with torsion:
\beq\label{sc1} 
D^{\pm}_{\mu}\,v_{\nu}=\nabla_{\mu}v_{\nu}\pm\frac 12\,T^{\si}_{\mu\nu}\,v_{\si},
\qq\quad D^{\pm}_{\alf}\,g_{\mu\nu}=0,\eeq
both compatible with the metric.

The spin connections $\Om^{\pm}$ and their structure equations are 
\beq\label{scdef}
D^{\pm}_{\mu}\,R^a_{\nu}=-\Om^{\pm\,a}_{~~~b\,\mu}\,R^b_{\nu},\qq\quad
dR^a+\Om^{\pm\,a}_{~~~s}\wedge R^s\pm\,T^a=0.\eeq
The explicit formula for $\Om^-$ is \footnote{To avoid confusion we change $\nabla_a$, defined 
by (\ref{nabla}), into $\wti{\pt}_a$ defined by $\pt_{\mu}=R^a_{\mu}\,\wti{\pt}_a$.}
\beq\label{conn}
\Om^-_{abc}=\frac 12\Big(\wti{\pt}_b\,G_{ac}+\wti{\pt}_c\,G_{ba}-\wti{\pt}_a\,G_{bc}\Big)
+\frac 12\Big(-f^s_{ab}\,G_{sc}+f^s_{ca}\,G_{bs}-f^s_{cb}\,G_{as}\Big), \eeq
Raising one index, we have 
\beq\label{om+}
\Om^{\pm a}_{~~~bc}=g^{as}\,\Om^{\pm}_{sbc},\qq\quad \Om^{+a}_{~~~bc}-\Om^{-a}_{~~~cb}=f^a_{bc},\eeq
as well as
\beq
\Om^{-\,s}_{~~~a,s}=\Om^{+\,s}_{~~~a,s}=\wti{\pt}_a\,\ln(\sqrt{\det g})+f^s_{as}.
\eeq
Working with the connection $\Om^-$, we will define the curvature as
\beq\label{curv} 
R^a_{~b}=d\Om^{-a}_{~~~b}+\Om^{-a}_{~~~s}\wedge \Om^{-s}_{~~~b}
=\frac 12\,R^a_{~b,st}\,R^s\wedge R^t,\eeq
which gives
\beq\label{riem}
R^a_{~b,cd}=\wti{\pt}_c\,\Om^{-a}_{~~~bd}-\wti{\pt}_d\,\Om^{-a}_{~~~bc}+\Om^{-a}_{~~~bs}f^s_{cd}+\Om^{-a}_{~~~sc}\Om^{-s}_{~~~bd}-\Om^{-a}_{~~~ud}\Om^{-u}_{~~~bc}.\eeq
Defining the Ricci tensor and scalar as \footnote{The sphere has positive curvature.}
\beq\label{ricdef}
Ric_{ab}=R^s_{~a,sb},\qq\qq R=g^{ab}\,Ric_{ab},
\eeq
and using relation (\ref{om+}) we end up with
\beq\label{ricci}
Ric_{ab}=\wti{\pt}_s\,\Om^{-s}_{~~~ab}-\Om^{-s}_{~~~a,t}\,\Om^{+t}_{~~~b,s}
-{\cal D}_b\,v_a,\qq v_a=\Om^{-s}_{~~~a,s},\eeq
using the frame covariant derivative
\beq\label{covder}
{\cal D}_b\,v_a=\wti{\pt}_b\,v_a-\Om^{-s}_{~~~ab}\,v_s.\eeq

\subsection{Metric and torsion potential}
As seen in Section 2 the classical action is
\beq
S=\int\,\Big(M+\Pi(g)\Big)^{-1}_{ab}\,R^a_+(g)\,R^b_-(g)\,d\xi^+d\xi^-.\eeq
Comparing this expression with (\ref{metric}) we can write
\beq 
G_{ab}=(M+\Pi)^{-1}_{ab}=g_{ab}+h_{ab},\qq \Ga^{ab}=M^{ab}+\Pi^{ab},\qq  G_{as}\,\Ga^{sb}=\Ga^{bs}\,G_{sa}=\de_a^b.\eeq
The riemannian metric can be written in two ways
\beq g=\frac 12(G+G^t)=G\,M_S\,G^t=G^t\,M_S\,G,\qq M_S=\frac 12(M+M^t),\eeq
as well as its inverse   
\beq
g^{-1}=\Ga^t\,M_S^{-1}\,\Ga=\Ga\,M_S^{-1}\,\Ga^t.\eeq

\subsection{The spin connection}
Let us write relation (\ref{conn}) as
\beq\label{conn1}
2\,\Om^-_{abc}=\Big(\wti{\pt}_b\,G_{ac}+f^s_{bc}\,G_{as}\Big)
+\Big(\wti{\pt}_c\,G_{ba}+f^s_{ca}\,G_{bs}\Big)-\Big(\wti{\pt}_a\,G_{bc}+f^t_{ab}\,G_{tc}\Big).\eeq
Using the identity (\ref{idPi}) we can combine, in the first term of (\ref{conn1}), the 
two pieces to obtain
\beq
\wti{\pt}_b\,G_{ac}+f^s_{bc}\,G_{as}=-G_{as}\,X^{st}_b\,G_{tc},\eeq
with 
\beq
X^{ab}_c=\wti{F}^{ab}_c-\Pi^{as}\,f^b_{sc},\qq \wti{F}^{ab}_c=\tilde{f}^{ab}_c-f^a_{bs}\,M^{sc}.\eeq
Similar computations for the remaining terms give 
\beq
2\,\Om^-_{abc}=-G_{as}\,X^{st}_b\,G_{tc}-G_{bs}\,X^{st}_c\,G_{ta}+G_{bs}\,Y^{st}_a\,G_{tc},
\eeq
with
\beq
Y^{ab}_c=\wh{F}^{ab}_c+f^a_{cs}\,\Pi^{sb},\qq \wh{F}^{ab}_c=\tilde{f}^{ab}_c+M^{as}\,f^b_{sc}.\eeq 

Raising the first index of the connection, we have
\beq
2\Om^{-a}_{~~~bc}=-g^{a\alf}G_{\alf s}\,X^{st}_b\,G_{tc}-G_{bs}\,X^{st}_c\,G_{t\alf}g^{\alf a}
+g^{a\alf}Y^{st}_{\alf}\,G_{bs}\,G_{tc}.\eeq
Using $g^{ab}=(\Ga\,M_S^{-1}\,\Ga^t)^{ab}$ one gets
\beq\label{conn2}
g^{a\alf}G_{\alf s}=2\de^a_s-(\Ga\,M_S^{-1})^a_{~s},\qq G_{t\alf}g^{\alf a}=(\Ga\,M_S^{-1})^a_{~t},\eeq
leading to
\beq
2\Om^{-a}_{~~~bc}=-2X^{as}_b\,G_{sc}
+(\Ga\,M_S^{-1})^a_{~v}\Big(X^{vt}_u\,\Ga^{su}-X^{sv}_u\,\Ga^{ut}+Y^{st}_u\,\Ga^{uv}\Big)G_{bs}\,G_{tc}.
\eeq
The second term in the right hand side is a polynomial of degree 2 in the $\Pi$'s. Getting 
rid of the quadratic terms upon use of the identity (\ref{idQ2}) all the terms linear 
in $\Pi$ cancel out and 
\beq\label{coone}
\Om^{-a}_{~~~bc}=-X^{as}_b\,G_{sc}+G_{bs}\,\Ga^{au}\,\wh{\cal F}^{st}_u\,G_{tc},\eeq
with
\beq\label{Fhat}
\wh{\cal F}^{st}_u=\frac 12\,(M_S^{-1})_{uv}
\Big(\wh{F}^{st}_a\,M^{av}+\wti{F}^{vt}_a\,M^{sa}-\wti{F}^{sv}_a\,M^{at}\Big).
\eeq

For further use, let us define 
\beq
\wti{\cal F}^{st}_u=\frac 12\,(M_S^{-1})_{uv}
\Big(\wti{F}^{st}_a\,M^{va}+\wh{F}^{sv}_a\,M^{at}-\wh{F}^{vt}_a\,M^{sa}\Big),
\eeq
and mention the identities
\beq\label{idconn}
\wh{F}^{ab}_c+\wti{F}^{ab}_c-\wh{\cal F}^{ab}_c-\wti{\cal F}^{ab}_c=\tilde{f}^{ab}_c,\qq 
\wti{\pt}_s\,\Pi^{ab}=X^{ab}_s+f^a_{su}\,\Ga^{ub}=Y^{ab}_s-\Ga^{au}\,f^b_{us}.\eeq
Using these relations and (\ref{om+}) and (\ref{coone}) we can write
\beq\label{cotwo}
\Om^{+a}_{~~~bc}=-\wti{\pt}_c\,\Pi^{as}\,G_{sb}+G_{cs}\,\Ga^{au}\,\wh{\cal F}^{st}_u\,G_{tb}.\eeq

If, instead of relations (\ref{conn2}), we use for the inverse metric its form $g^{ab}=(\Ga^t\,M_S^{-1}\,\Ga)^{ab}$, then instead of (\ref{conn2}) we have
\beq
g^{a\alf}\,G_{\alf s}=(\Ga^t\,M_S^{-1})^a_{~s},\qq G_{t\alf}\,g^{\alf a}=2\de^a_t-(\Ga^t\,M_S^{-1})^a_{~t},
\eeq
leading this time to
\beq
2\Om^{-a}_{~~~b,c}=-2X^{sa}_c\,G_{bs}
+(\Ga^t\,M_S^{-1})^a_{~v}\Big(Y^{st}_u\,\Ga^{vu}-X^{vt}_u\,\Ga^{su}+X^{sv}_u\,\Ga^{ut}\Big)G_{bs}\,G_{tc}.
\eeq
Computations similar to the ones leading to (\ref{coone}) give then
\beq\label{cothree}
\Om^{-a}_{~~~b,c}=-X^{sa}_c\,G_{bs}+G_{bs}\,(\Ga^t)^{au}\Big(Y^{st}_u-\wh{\cal F}^{st}_u\Big)\,G_{tc}.\eeq

\subsection{The Ricci tensor}\label{sric}
Let us start from (\ref{ricci}):
\[Ric_{ab}=\wti{\pt}_s\,\Om^{-s}_{~~~ab}-\Om^{-s}_{~~~a,t}\,\Om^{+t}_{~~~b,s}
-{\cal D}_b\,v_a,\qq v_a=\wti{\pt}_a(\ln\sqrt{\det g})+f^s_{as}.\]
In the first term we use (\ref{coone}), while in the product we use (\ref{cothree}) for $\Om^-$ and 
(\ref{cotwo}) for $\Om^+$. The resulting expression is either cubic or quadratic in $G$. We will write it as
\beq
Ric_{ab}=G_{a\alf}\,{\cal L}^{\alf\be}\,G_{\be b}
+G_{a\alf}\,{\cal M}^{\alf\be,\la\mu}\,G_{\la\mu}\,G_{\be b}.\eeq
The quantities ${\cal L}$ (resp. ${\cal M}$) are quadratic (resp. cubic) with respect to $\Pi$. Let us first  explain how one can get rid of the terms cubic in $G$. Let us consider:
\beq\label{eqric1}
{\cal M}^{\alf\be,\la\mu}\,G_{\la\mu}=(\Ga^{\alf s}\,X_s^{t\la}+Y_s^{\alf\la}\,\Ga^{st}
-2\wh{\cal F}_s^{\alf\la}\,M_S^{st})\,G_{\la\mu}\wti{\pt}_t\,\Pi^{\mu\be}
\equiv H^{\alf t \la}\,G_{\la\mu}\wti{\pt}_t\,\Pi^{\mu\be}.\eeq
The terms in $H$ are quadratic in $\Pi$ and can be reduced as follows. Using the identity  (\ref{idQ2}) the terms quadratic in $\Pi$ reduce to
\beq
f^t_{uv}\,\Pi^{\alf v}\,\Pi^{u\la}=f^t_{uv}\,\Pi^{\alf v}\,\Ga^{u\la}
-f^t_{uv}\,\Pi^{\alf v}\,M^{u\la},\eeq
and the first piece, when multiplied by $\,G_{\la\mu}\,$, gives a contribution quadratic in G. Then the terms remaining in $H$ are just linear in $\Pi$ and they combine to
\beq
\wti{F}^{\alf t}_s\,\Pi^{s\la}=\wti{F}^{\alf t}_s\,\Ga^{s\la}
-\wti{F}^{\alf t}_s\,M^{s\la}.\eeq
The first piece gives another contribution quadratic in G. The remaining terms in $H$ 
are now independent of $\Pi$ and vanish as a consequence of (\ref{Fhat}). 

Gathering all of the pieces quadratic in $G$ we get, apart from a few trivial cancellations
\beq
\wti{\pt}_s\,\Om^{-s}_{~~~ab}-\Om^{-s}_{~~~a,t}\,\Om^{+t}_{~~~b,s}=G_{a\alf}\Big\{(f^{\be}_{tu}\,\Ga^{\alf u}+\wh{\cal F}^{\alf\be}_t)\wti{\pt}_s\Pi^{st}
-(Y^{\alf s}_t-\wh{\cal F}^{t\be}_s+f^{\alf}_{ut} \Ga^{us})\wh{\cal F}^{t\be}_s\Big\}G_{\be b}.\eeq
The second term, in the right hand part simplifies, using (\ref{idconn}), to
\beq\label{ric1}
\wti{\pt}_s\,\Om^{-s}_{~~~ab}-\Om^{-s}_{~~~a,t}\,\Om^{+t}_{~~~b,s}=G_{a\alf}\Big\{(f^{\be}_{tu}\,\Ga^{\alf u}+\wh{\cal F}^{\alf\be}_t)\wti{\pt}_s\Pi^{st}
+\wti{\cal F}^{\alf s}_t\,\wh{\cal F}^{t\be}_s\Big\}G_{\be b}.\eeq

To get rid of the residual $\Pi$ dependence, we will consider the vector field 
$\,w_a=G_{as}\,\wti{\pt}_t\Pi^{ts}$. Using (\ref{covder}) and (\ref{cothree}) we get
\beq\label{res}
{\cal D}_b\,w_a=G_{a\alf}\Big\{(\wti{\pt}^{\,2}_{st}\Pi^{t\alf}
-f^{\alf}_{su}\,\wti{\pt}_t\Pi^{tu})\Ga^{s\be}
-(Y^{\alf\be}_s-\wh{\cal F}^{\alf\be}_s)\wti{\pt}_t\Pi^{ts}\Big\}G_{\be b}.\eeq
The relation (\ref{idspe1}), proved in appendix A, shows that the coefficient of 
$\Ga^{s\be}$ in the previous relation does vanish. So we can use (\ref{res}) in the 
following way
\beq
G_{a\alf}\,\wh{\cal F}^{\alf\be}_s\,\wti{\pt}_t\Pi^{ts}\,G_{\be b}={\cal D}_b\,w_a
+G_{a\alf}\,\wti{\pt}_t\,\Pi^{ts}\,Y^{\alf\be}_s\,G_{\be b},
\eeq
to transform (\ref{ric1}) into
\beq
\wti{\pt}_s\,\Om^{-s}_{~~~ab}-\Om^{-s}_{~~~a,t}\,\Om^{+t}_{~~~b,s}={\cal D}_b\,w_a+
G_{a\alf}\Big\{\wti{\cal F}^{\alf s}_t\,\wh{\cal F}^{t\be}_s
+\wti{\pt}_s\Pi^{st}\,\wti{\pt}_t\Pi^{\alf\be}\Big\}G_{\be b}.\eeq
The residual term quadratic in the derivatives of $\Pi$ does vanish as a consequence of the relation (\ref{idspe2}) 
proved in appendix A. Therefore we have 
\beq\label{ricfin1}\left\{\barr{l}
Ric_{ab}=G_{a\alf}\,r_0^{\alf\be}\,G_{\be b}+{\cal D}_b\,(w_a-v_a),\\[4mm] 
r_0^{\alf\be}=\wti{\cal F}^{\alf s}_t\,\wh{\cal F}^{t\be}_s,\\[4mm]
v_a=\wti{\pt}_a(\ln\sqrt{\det g})+f^s_{as},\qq w_a=G_{as}\,\wti{\pt}_t\Pi^{ts}.\earr\right.\eeq

Of course the Ricci tensor is uniquely defined, however the {\em writing} of the result is not unique 
for the following reason. Let us consider a vector of the form $\,W_a=G_{a\alf}\,\xi^{\alf}$. 
Using the connection given by (\ref{cothree}), we get
\beq
{\cal D}_b\,W_a=\Big(\wti{\pt}_s\,\xi^{\alf}\,\Ga^{s\be}
-\xi^s(f^{\alf}_{us}\,\Ga^{u\be}+Y^{\alf\be}_s
-\wh{\cal F}^{\alf\be}_s)\Big)G_{a\alf}\,G_{\be b}.\eeq
The second term simplifies and if we consider coordinate independent $\xi^{\alf}$ we get 
\beq
{\cal D}_b\,W_a=-\xi^s\,\wti{\cal F}^{\alf\be}_s\,G_{a\alf}\,G_{\be b}.\eeq
This generates an ambiguity since we can write 
\beq\label{ricfin2}
Ric_{ab}=G_{a\alf}\,r^{\alf\be}\,G_{\be b}+{\cal D}_b\,(w_a+W_a-v_a),\qq 
r^{\alf\be}=r_0^{\alf\be}+\xi^t\,\wti{\cal F}^{\alf\be}_t,\eeq
for any coordinate independent $\xi^{\alf}$. 
>From the previous result, by the obvious duality substitutions, one can easily get the Ricci tensor for the dual model.

\section{One loop renormalizability}
We started from the classical action 
\beq
S=\frac 1{2}\,\int\,G_{ab}\,R^a_+\,R^b_-\ d\xi^+ d\xi^-.\eeq
The one loop counterterm was first computed by Fridling and van de Ven  \cite{FV} 
for $\si$-models with general torsion. Their result is
\beq
\frac 1{4\pi\eps}\int\,Ric_{ab}\,R^a_+\,R^b_-\ d\xi^+ d\xi^-,\qq\quad \eps=2-d,\eeq
where the Ricci tensor is computed with the $D^-$ connection. 

Renormalizability in the strict field theoretic sense requires that these divergences have 
to be absorbed by {\em field independent} deformations of the coupling constants $\,M^{st}$ and possibly a non-linear field renormalization of the fields $X^{\mu}$. We have first to define the coupling constants of the theory. Since we are analyzing the most general Poisson-Lie model, 
which is built up from an arbitrary constant matrix $M$, we will take as independent coupling constants all the matrix elements of $M$.  When working in coordinates, we have 
to check the relations
\beq
Ric_{(\mu\nu)}=\chi^{st}\frac{\pt}{\pt M^{st}}\,g_{\mu\nu}+\nabla_{(\mu}u_{\nu)},\qq
Ric_{[\mu\nu]}=
\chi^{st}\frac{\pt}{\pt M^{st}}\,h_{\mu\nu}+T_{\mu\nu}^{\si}\,u_{\si}+\pt_{[\mu}U_{\nu]},
\eeq
for some vectors $u,\,U$ and coordinate independent $\chi^{st}$. Adding these relations we can write
\beq
Ric_{\mu\nu}=\chi^{st}\frac{\pt}{\pt M^{st}}\,G_{\mu\nu}+D^-_{\nu}\,u_{\mu}+\pt_{[\mu}(u+U)_{\nu]},
\eeq
and since the frames are independent of the parameters of the matrix $M$, this relation becomes, 
using frame components
\beq
Ric_{ab}=\chi^{st}\frac{\pt}{\pt M^{st}}\,G_{ab}+{\cal D}_b\,u_a+\wti{\pt}_{[a}(u+U)_{b]}
+f^s_{ab}\,(u+U)_s.\eeq
Since relation (\ref{ricfin2}) writes
\beq
Ric_{ab}=G_{a\alf}\,r^{\alf\be}\,G_{\be b}+{\cal D}_b\,(w_a+W_a-v_a),\eeq
we obtain 
\beq
\chi^{ab}=-r^{ab},\qq U_a=-u_a,\qq u_a=w_a+W_a-v_a,
\eeq
and this concludes the one loop renormalizability proof. 

Let us observe that the non-uniqueness of $r^{ab}$ explained in the previous section has for 
effect that the renormalization constants themselves are not uniquely defined. This phenomenon had 
already been observed in \cite{KV}.

\appendix

\section{Some identities}
We have gathered in this Appendix the proofs of various identities used in the article.
\subsection{Derivative of the bivector}
Let us start from the relation which defines the matrices $\,A\,$ and $\,B\,$:
\beq\label{defAB}
g\,\Tt^a\,g^{-1}=B^{sa}(g)\,T_s+A^a_s(g)\,\Tt^s.\eeq
Differentiating both sides, expressing $\,dg$ as $R^a\,T_a\,g$ and using the commutation 
relations for the Drinfeld double gives
\beq
dB^{ab}(g)=R^s(g)\Big(f^a_{st}\,B^{tb}-\tf^{at}_s\,A^b_t(g)\Big),\qq 
dA^a_b(g)=-R^s(g)\,A^a_t(g)\,f^t_{sb}.\eeq
Recalling the definition of the directional derivative
\beq
\wti{\pt}_s\,f(g)=\frac d{ds}\,f(e^{sT^a}g)\Big|_{s=0}=R^s_a(g)\frac{\pt}{\pt x^s}f(g),\qq 
g=e^{x^sT_s}\in G,\eeq
we end up with
\beq\label{RA}\left\{\barr{l}
\wti{\pt}_s\,B^{ab}(g)=f^a_{st}\,B^{tb}(g)-\tf^{at}_s\,A^b_t(g),\\[4mm]  
\wti{\pt}_s\,A^a_b(g)=-A^a_t(g)\,f^t_{sb},\qq   \wti{\pt}_s\,A^a_b(g^{-1})=f^a_{st}\,A^t_b(g^{-1}).\earr\right.\eeq                     
Recalling the definition of the bivector $\ \Pi^{ab}(g)=-B^{as}(g)\,A^b_s(g^{-1}),\,$ 
we conclude to 
\beq\label{derpi}
\wti{\pt}_s\,\Pi^{ab}(g)=\tilde{f}^{ab}_s-f^a_{st}\,\Pi^{bt}(g)+f^b_{st}\,\Pi^{at}(g).
\eeq 
This derivative can also be interpreted as
\beq\label{ftig}
\wti{\pt}_c\,\Pi^{ab}(g)=\tf^{ab}_c(g)\equiv 
A^a_s(g^{-1})\,A^b_t(g^{-1})\,\tf^{st}_u\,A^u_c(g).
\eeq
To prove this relation, let us start from 
\[
\tf^{st}_u=\langle [\Tt^s,\Tt^t],T_u\rangle\quad\Rightarrow\quad 
\tf^{ab}_c(g)=\langle [A^a_s(g^{-1})\,\Tt^s,A^b_t(g^{-1})\,\Tt^t],A^u_c(g)\,T_u\rangle.\]
Using (\ref{defAB}) we can write
\[
\tf^{ab}_c(g)=\langle[g^{-1}\Tt^a g-B^{au}(g)\,T_u,g^{-1}\Tt^b g-B^{bv}(g)\,T_v],
g^{-1}\Tt^a g\rangle.\]
Due to the isotropy property only three terms out of four do not vanish. The $Ad_g$ invariance 
gives for the first term
\[
\langle g^{-1}[\Tt^a,\Tt^b]g,g^{-1}T_s g\rangle=\langle [\Tt^a,\Tt^b],T_s \rangle=\tf^{ab}_c.\]
The second term is
\[\barr{l}
=-B^{au}(g)\langle [T_u,g^{-1}\Tt^b g], g^{-1} T_s g\rangle=
-B^{au}(g)\langle [g^{-1}T_u g,\Tt^b ], T_s \rangle= \\[4mm]
=-B^{au}(g)\,A_u^v(g^{-1})\langle [T_v ,\Tt^b ], T_s \rangle=\Pi^{av}(g)\,(-f^b_{vs})
=f^b_{st}\,\Pi^{at}(g),\earr\]
and the third term is just the opposite of the second term with the exchange $(a\leftrightarrow b)$, 
and this concludes the proof.

\subsection{Identities involving $\Pi$ derivatives}
The transition from $\tf^{ab}_c$ to $\tf^{ab}_c(g)$ amounts to a change of basis in the 
Lie algebra $\,{\cal G}$. It follows that they do verify the Jacobi identity:
\beq
\tf^{st}_c(g)\,\tf^{ab}_t(g)+\tf^{at}_c(g)\,\tf^{bs}_t(g)+\tf^{bt}_c(g)\,\tf^{sa}_t(g)=0.\eeq
Contracting the indices $c$ and $s$ we get
\beq\label{jacspe}
\tf^{st}_s(g)\,\tf^{ab}_t(g)=0,\eeq
which becomes, using relation (\ref{ftig}) 
\beq\label{idspe2}
\wti{\pt}_s\,\Pi^{st}(g)\,\wti{\pt}_t\,\Pi^{ab}(g)=0.\eeq

Let us notice that
\[
\tf^{sa}_s(g)=\tf^{st}_s\,A_t^a(g^{-1}).\]
Using the last relation in (\ref{RA}) we get
\[
\wti{\pt}_t\,\tf^{sa}_s(g)=\tf^{st}_s\,\,f^a_{tu}\,A^u_t(g^{-1})=f^a_{tu}\,\tf^{su}_s(g),\]
and from (\ref{ftig}) we conclude to the identity
\beq\label{idspe1}
\wti{\pt}_t\,\wti{\pt}_s\,\Pi^{sa}(g)=f^a_{tu}\,\wti{\pt}_s\Pi^{su}(g).\eeq

\end{document}